\documentstyle[12pt]{article}

\def\tr{{\rm Tr}}
\def\a{\begin{eqnarray}}
\def\b{\end{eqnarray}}
\def\0{\nonumber}
\def\ba{\begin{array}}
\def\ea{\end{array}}

\newlength{\extraspace}
\newlength{\extraspaces}
\newcounter{dummy}

\newcommand{\ai}{
\addtocounter{equation}{1}
\setcounter{dummy}{\value{equation}}
\setcounter{equation}{0}
\renewcommand{\theequation}{\thesection.\arabic{dummy}\alph{equation}}
\begin{eqnarray}
\addtolength{\abovedisplayskip}{\extraspaces}
\addtolength{\belowdisplayskip}{\extraspaces}
\addtolength{\abovedisplayshortskip}{\extraspace}
\addtolength{\belowdisplayshortskip}{\extraspace}}
\newcommand{\bj}{
\end{eqnarray}
\setcounter{equation}{\value{dummy}}
\renewcommand{\theequation}{\thesection.\arabic{equation}}}

\def\d{{\partial}}

\newcommand{\bac}{\begin{array}{c}}
\newcommand{\bacc}{\begin{array}{cc}}
\newcommand{\baccc}{\begin{array}{ccc}}
\newcommand{\barcl}{\begin{array}{rcl}}
\newcommand{\bacccc}{\begin{array}{cccc}}
\newcommand{\baccccc}{\begin{array}{ccccc}}
\newcommand{\baccccccc}{\begin{array}{ccccccc}}
\newcommand{\barclcrcl}{\begin{array}{rclcrcl}}
\newcommand{\bacl}{\begin{array}{cl}}
\newcommand{\bal}{\begin{array}{l}}
\newcommand{\bacll}{\begin{array}{cll}}

\def\cs{{\cal S}}
\def\cas{{\cal AS}}

\begin{document}
\vskip0.5cm

\centerline{\LARGE\bf Remark on Quantum Nambu Bracket}

\vskip1cm
\centerline{\large Chuan-Sheng Xiong}
\vskip0.5cm
\centerline{Department of Physics, Beijing University}
\centerline{Beijing 100871, China}
\centerline{email: xiong@ibm320h.phy.pku.edu.cn}
\vskip1.5cm

\centerline{\bf Abstract}

We give an explicit realization of quantum Nambu bracket
via matrix of multi-index, which reduces in the continunm 
limit to the classical Nambu bracket.

\vskip1cm

\section{Introduction}

In the original paper \cite{Nambu} Nambu introduced his famous
Poisson bracket, which describes a generalized Hamilton mechanics. 
Let $(x^i,i=1,2,\cdots, n)$ be a set of dynamical variables 
which span a $n$-dimensional phase space. 
A Nambu-Poisson structure is defined as 
\a
\{f_1, f_2,\cdots, f_n\}:=\sum_{i_1,i_2,\cdots,i_n}
\epsilon_{i_1,i_2,\cdots,i_n} \frac{\d f_1}{\d x_{i_1}}
\frac{\d f_2}{\d x_{i_2}}\cdots \frac{\d f_n}{\d x_{i_n}},
\label{nambupb}
\b
where $\epsilon_{i_1,i_2,\cdots,i_n}$ is the $n$-dimensional
Levi-Civita tensor. This bracket has the following properties
\cite{Tak}
\begin{enumerate}
\item
{\it Skew-symmetry}\,:
\a
\{f_1, f_2,\cdots, f_n\}=(-1)^{\epsilon(p)}
\{f_{p(1)}, f_{p(2)},\cdots, f_{p(n)}\},\0
\b
where $p(i)$ is the permutation of indices and $\epsilon(p)$ is
the parity of the permutation.
\item
{\it Derivation}\,:
\a
\{f_1 f_2,f_3,\cdots, f_{n+1}\}=f_1 \{f_2,f_3,\cdots, f_{n+1}\}+
\{f_1 ,f_3,\cdots, f_{n+1}\}f_2.\0
\b
\item
{\it Fundamental Identity}\,\,(FI):
\a
&&\qquad\{\{f_1,f_2,\cdots, f_n\},f_{n+1},\cdots,f_{2n-1}\}=
\{\{f_1,f_{n+1},\cdots,f_{2n-1}\},f_2,\cdots, f_n\}
\0\\
&&+\{f_1,\{f_2, f_{n+1},\cdots,f_{2n-1}\},f_3,\cdots, f_n\}
+\cdots+\{f_1,\cdots, f_{n-1},\{f_n,f_{n+1},\cdots,f_{2n-1}\}\},\0
\b
which is a generalization of the Jacobi identity.
\end{enumerate}
The dynamics of a Nambu system is determined by $(n-1)$ Hamiltonians
$H_1,H_2,\cdots,H_{n-1}$ and is described by 
Nambu-Hamiltonian equation
\a
\frac{d f}{d t}=\{f, H_1,H_2,\cdots,H_{n-1}\}.
\b

The quantization of Nambu bracket turns out to be a quite 
non-trivial problem. In general one expects that the quantum
Nambu bracket satisfies some properties analogous to those of
the classical Nambu bracket, i.e.
\begin{enumerate}
\item
{\it Skew-symmetry},
\a
[A_1, A_2,\cdots, A_n]=(-1)^{\epsilon(p)}
[A_{p(1)}, A_{p(2)},\cdots, A_{p(n)}].
\b
where $p(i)$ is the permutation of indices and $\epsilon(p)$ is
the parity of the permutation.
\item
{\it Derivation}\,:
\a
[A_1 A_2,A_3,\cdots, A_{n+1}]=A_1 [A_2,A_3,\cdots, A_{n+1}]+
[A_1 ,A_3,\cdots, A_{n+1}]A_2.\label{qd}
\b
\item
{\it Fundamental Identity}\,\,(FI):
\a
&&\qquad [[A_1,A_2,\cdots, A_n],A_{n+1},\cdots,A_{2n-1}]=
[[A_1,A_{n+1},\cdots,A_{2n-1}],A_2,\cdots, A_n]
\label{qFI}\\
&&+[A_1,[A_2, A_{n+1},\cdots,A_{2n-1}],A_3,\cdots, A_n]
+\cdots+[A_1,\cdots,A_{n-1},[ A_n,A_{n+1},\cdots,A_{2n-1}]].\0
\b
\end{enumerate}
To maintain the skew-symmetry, it is natural to 
introduce the following quantum commutator\cite{Nambu}
\a
[A_1,A_2,\cdots,A_n]:=\sum_{i_1,i_2,\cdots,i_n}
\epsilon_{i_1,i_2,\cdots,i_n}(A_{i_1}A_{i_2}\cdots A_{i_n}).
\label{qcgen}
\b
Thus the problem of quantizing Nambu bracket reduces to 
finding a proper quantum product, which maps $n$ objects 
to one, such that the quantum commutator (\ref{qcgen})
satisfies the derivation law (\ref{qd}) and the fundamental
identity (\ref{qFI}). This turns out to be a very 
difficult problem. Several quantization schemes have been 
proposed in the literatures\cite{Dito}\cite{Flato}\cite{Awata}.
Among them, the Zariski quantization is a deformation quantization,
which obeys all the three requirements listed in the above,
and has a clear relationship with the classical Nambu bracket
\cite{Dito}. In the other approaches, the quantum Nambu bracket 
does not satisfy the derivation law \cite{Flato}\cite{Awata}. 
In particular, Awata etc. proposed a matrix model realization of
the quantum Nambu bracket\cite{Awata}, but its link to
the classical Nambu bracket is quite obscure.
We will give another matrix realization of the quantum 
Nambu bracket, which redues in the continuum limit  
to the classical Nambu bracket.

\section{Quantum Nambu bracket}

Let each of $\{A_i, i=1,2,\cdots,n\}$ denote a
${\overbrace{N\times N\times\cdots\times N}^n}$ matrix,
and $\cs_n(N)$ the collection of all such matrices.
we define the quantum product as follows
\footnote{Throughout the paper, any summation index 
will be specified explicitly.}
\a
(A^{(1)}A^{(2)}\cdots A^{(n)})_{i_1 i_2,\cdots,i_{n}}
:=\sum_l A^{(1)}_{i_1 i_2\cdots i_{n-1}l}\cdots
 A^{(n-1)}_{i_1 l i_3\cdots  i_n} 
 A^{(n)}_{l i_2\cdots i_n}.\label{progen}
\b
We may also define the inner scalar product, i.e.
the ``{\it Trace}"
\a
\tr(A):=\sum_{i=1}^N A_{i i\cdots i}.
\b
When $n=2$, we recover the usual product of the square
matrix. When $n\geq3$, the quantum product (\ref{progen})
is {\it non-commutative}\, and {\it non-associative}. The
derivation law (\ref{qd}) does 
not hold since the product of two objects $A_1$ and $A_2$
does not make sense. Even worse, the above quantum product 
generally does not make the quantum commutator (\ref{qcgen}) 
satisfying the Fundamental Identity (\ref{qFI}).
We would expect that there exists some subset 
of $\cs_n(N)$, which is closed under
the quantum commuator (\ref{qcgen}), and
obeys on the Fundamental Identity (\ref{qFI}).
We will give some examples to show that this is indeed 
the case.

In this article we only consider the quantization of the 
the even-dimensional Nambu bracket. To do so,
set $n=2m(m\geq2)$. Let $\cas_{2m}(N)$ denote the
subset of $\cs_{2m}(N)$ containing all matrices 
which are totally antisymmetric, i.e. 
\a
A_{i_1 i_2\cdots  i_{2m}}=(-1)^{\epsilon(p)}
A_{p(i_1) p(i_2)\cdots p(i_{2m})}.\0
\b
It is obvious to see that this trucation is consistent 
with the quantum commutator(\ref{qcgen}), if we use 
the quantum product (\ref{progen}). When $N<2m$, the
subset $\cas_n(N)$ is empty. When $N=2m$, $\cas_n(N)$ contains
just one element, which generates an abelian algebra.
So the Fundamental Identity is trivial. The first 
non-trivial example is the case $N=2m+1$, where $\cas_{2m}(N)$
contains $N$ independent elements, which we may 
denote by
\a
(T_a)_{i_1i_2\cdots i_{2m}}:=
\epsilon_{a i_1i_2\cdots i_{2m}},\qquad a=1,2,\cdots,N.
\label{egenerators}
\b
It is straightforward to show that these matrices generate
a quantum algebra with the following commutation relation
\a
[T_{a_1}, T_{a_2}, \cdots, T_{a_{2m}}]
=\sum_b\epsilon_{a_1a_2\cdots a_{2m}b}T_b.
\label{ecom}
\b
The fundamental identity is guaranteed by the following
equality
\a
\sum_c\Bigl(\epsilon_{c a_1a_2\cdots a_n}
\epsilon_{c b_1 b_2\cdots b_n}
-\sum_{i=1}^n(a_i\leftrightarrow b_n)\Bigl)=0.
\label{Levi-Civita}
\b
One may generalize the consideration here to the case with 
the general value of $N$.

\section{The continuum limit}

The quantum product (\ref{progen}) has a very simple geometric 
meaning. Consider  $N(N\geq n)$ ordered points 
in $(n-1)$-dimensional Euclidean space ${\bf R}^{n-1}$.
For each $(n-1)$-dimensional convex with vertices  
$(i_1,i_2,\cdots, i_n)$ in general positions, we associate 
to it a matrix element $A_{i_1,i_2,\cdots, i_n}$.
Adding a point $l$ to this convex leads to $n$ more 
convexes, all of which contain the point $l$. Thus we
have a natural composition law,
\a
\underbrace{\cs\times \cs\times \cdots\times \cs}_n
\longrightarrow \cs,\qquad (A_{i_1 i_2,\cdots,i_n})\in \cs,
\b
which glues $n$ convexes to one, and defines a general 
``{\it group structure}". The explicit form of this composition 
law is given by the quantum product (\ref{progen}).

The geometric picture provides a way  to analyse the
continuum limit of the quantum Nambu bracket. To do so,
we may take the following viewpoint. Each matrix defines a
map from the set of convexes to the field ${\bf R}$ 
or ${\bf C}$,
\a
A:\{{\rm convexes}\}\longrightarrow {\bf R}({\rm or}\, {\bf C}).\0
\b
Each convex can be characterized by its center and volume
\footnote{We may think of that the ordered points are sited
on a $(n-1)$-dimensional rectangular lattice. The 
volume-preserving symmetry group of the lattice is  
$sl(n-1,{\bf Z})$. We are interested in the 
quotient lattice space, which is the regular lattice
modulo out $sl(n-1,{\bf Z})$ symmetry.},
which is 
\a
&&{\vec x}:=\frac{\epsilon}{n}({\vec i}_1
+{\vec i}_2+\cdots+{\vec i}_n)=(x_1,x_2,\cdots, x_{n-1}),
\qquad \epsilon:=\frac{1}{N},\0\\
&&{\rm vol}:=\frac{1}{(n-1)!}\epsilon_{\mu_1 \mu_2\cdots \mu_{n-1}}
({\vec i}_1-{\vec i}_n)_{\mu_1}
({\vec i}_2-{\vec i}_n)_{\mu_2}\cdots
({\vec i}_{n-1}-{\vec i}_n)_{\mu_{n-1}}, \0\\
&&{\vec v}^{(j)}={\vec i}_{n-j+1}-{\vec l}, \qquad 1\leq j\leq n.\0
\b
Where $(\mu_1,\mu_2,\cdots,\mu_{n-1})$ are indices of the Cartesian
coordinates in the $(n-1)$-dimensional space. To go to the continuum 
limit, we let all the $n$-indices $(i_1 i_2,\cdots,i_n)$
go to infinity, but their differences finite. Thus ${\vec x}$
become the continuum variables. To characterize the 
dependence of a matrix on the volume of the convex, we introduce
a new parameter $x_n$, such that
\a
\frac{\d f({\vec x}, x_n)}{\d x_n}:=f({\vec x}, x_n)\cdot {\rm vol},\0
\b
here $f({\vec x}, x_n)$ is the continuum limit of the matrix 
$A_{i_1 i_2,\cdots,i_n}$. One may think of $x_n$ as the constant 
density of mass, momentum or charge, etc. With these notations, 
we may expand the commutator (\ref{qcgen}) in the powers of $\epsilon$, 
and we find that
\a
[A^{(1)}, A^{(2)}, \cdots, A^{(n)}]=
(\frac{n!\epsilon^{n-1}}{n^{n-1}})
\{f^{(1)},f^{(2)},\cdots,f^{(n)}\} 
+{\cal O}(\epsilon^{n}).
\b

\section{Conclusion}

In this letter we have constructed a quantum Nambu bracket
via matrices of multi-index. We have shown that our quantum
commutator has the right continuum limit. 
Our construction may be helpful to quantize $p$-brane theory, 
just like what we did in studying $2$-dimensional quantum 
gravity by the ordinary matrix models. Furthermore if we regard 
the points in ${\bf R}^{n-1}$ as a set of $D0$-particles, 
we may expect a matrix theory in more general
setting. We wish to discuss these problems in more details
elsewhere.

\vskip0.8cm
\noindent
{\bf Acknowledgements}

We thank Drs. H. Awata, M. Li, X.C. Song, M. YU and C. J. Zhu
for discussions and comments.
The work is supported in part by NSFC grant 19925521 and by
the Startup grant from Beijing University.

\end{document}